\documentclass[12pt]{article}
\usepackage[dvips]{color}
\usepackage{epsfig}
\usepackage{amsmath}
\usepackage{graphicx}

\begin{document}

\title{Logarithmic corrections to charged hairy black hole in (2+1) dimensions}
\author{{J. Sadeghi $^{a}$\thanks{Email: pouriya@ipm.ir}\hspace{1mm}, B. Pourhassan  $^{b}$\thanks{Email:
b.pourhassan@du.ac.ir}\hspace{1mm},
and F. Rahimi  $^{c}$\thanks{Email: f.rahimi@umz.ac.ir}}\\
$^{a}$ {\small {\em  Department of Physics, Ayatollah Amoli Branch, Islamic Azad University,}}\\
{\small {\em P .O .Box 678, Amol, Iran}}\\
$^{b}${\small {\em School of Physics, Damghan University,
Damghan, Iran}}\\
$^{d}$ {\small {\em  Physics department, Mazandaran University, Babolsar, Iran}}} \maketitle
\begin{abstract}
\noindent\\
We consider a charged black hole with a scalar field that is coupled to gravity in (2 + 1)-dimensions. We compute the
logarithmic corrections to the corresponding system using two approaches. In the first method we take advantage of thermodynamic
properties. In the second method we use the metric function that is suggested by conformal field theory. Finally, we
compare the results of the two approaches.
\end{abstract}

\section{Introduction}
Gravity in $(2 + 1)$-dimensional space-time has been an interesting
area of theoretical researches during recent decades. Such studies
began in the early 1980 [1-4]. By  the discovery of
Ba˜nados{-} Teitelboim{-} Zanelli (BTZ) [5] and Martinez{-}
Teitelboim{-}Zanelli (MTZ) [6] black holes, it became clear
that gravity in $(2 + 1)$ dimensions is much more fascinating in its
own place and it is often much easier to analyze black hole
solutions in $(2 + 1)$ dimensions than it in other dimensions. Recently,
Xu and Zhao analyzed the charged black hole with scalar field in
$(2+1)$ dimensions, where the scalar field couples to gravity and
it couples to itself with the self-interacting potential too
[7]. Then, similar black hole with a rotational parameter constructed by the Ref. [8] and developed by the Ref. [9]. Also some thermodynamical studies of such kind of black hole may be find in the Refs. [10] and [11].\\
It is shown that the entropy of large black holes is
adequate to its horizon area [12, 13], so while one
reduces the size of the black hole, it is important to study what
the leading order corrections are, and also proven that when small
stable fluctuations around equilibrium are considered, logarithmic
corrections to thermodynamic entropy appear in all thermodynamic
systems. The stability condition is correspond to being positive of
the specific heat, so that the equivalent canonical ensemble is
stable [14]. We want to study this logarithmic corrections for
the charged black hole with scalar hair in three dimensions and the
special cases in section 3. We will study the thermodynamics of
these black holes that similarly was done in [11] and use them
for compute these corrections. in next section we try thermodynamic
entropy as an exact function of the inverse temperature $\beta (S =
S(\beta))$. In this part concentrate on the function offered by
CFT [15], leads to identical logarithmic corrections like
found in the earlier section from thermodynamic considerations.
Finally in section 6 we give conclusion.
\section{Charged hairy black hole in three dimensions and entropy}
The black hole solution in an Einstein{-}Maxwell scalar gravity that
coupled with a nonminimally scalar field in $(2 + 1)$ dimensions was
studied. Hairy black hole is the name that black hole solutions in
this theories have, and now there is lots of texts about this entry
[16-29], and the space-time is not only limited to be $(2
+ 1)${-}dimensional [30]. The scalar field may be
coupled minimally or nonminimally, to gravity, and it may coupled to
itself by a self{-}interacting potential $U(\varphi)$. In the model
that was studied, $\varphi$ couples to gravity in a nonminimal way,
and it also couples to itself by a self{-}potential $V (\varphi)$.
The action is given by,
\begin{equation}\label{m1}
s=\int{ d^3x \sqrt{-g}(\frac{R}{2} - \frac{1}{2} g^{\alpha\beta}
\nabla_{\alpha}\varphi \nabla_{\beta}\varphi - \frac{1}{2}
{\varepsilon} R \varphi^2 - V(\varphi)-\frac{1}{8} F_{\alpha\beta}
F^{\alpha\beta})},
  \end{equation}
where $\varepsilon=\frac{1}{8}$ is a constant shows the coupling power
between gravity and the scalar field and the metric
function is obtained as the following,
\begin{equation}\label{m2}
f(r)=\frac{r^2}{l^2} +3\beta-\frac{Q^2}{4}+(2\beta-\frac{Q^2}{9})
\frac{B}{r}-Q^2 (\frac{1}{2}+\frac{B}{3r}) \ln(r),
  \end{equation}
where $Q$ is the electric charge, $l$ and $\beta$ are integration
constants. $l$ related to the cosmological constant by
$\Lambda=-\frac{1}{l^2}$, it is negative because smooth black hole
horizons can be only in presence of a negative cosmological
constant in $(2+1)$ dimensions, and $r$ explains the radial
coordinate. the relation between $B$ and scalar field is as follow,
\begin{equation}\label{m3}
\phi(r)=\pm\sqrt{\frac{8B}{r+B}}.
  \end{equation}
Also,
\begin{equation}\label{m4}
\beta =\frac{1}{3}(\frac{Q^2}{4}-M),
  \end{equation}
  is a relation between the black hole charge and mass.
So we use them to calculate the related entropy as will see in
the next sections. 

\section{Logarithmic correction with thermodynamic
properties}
There is the exact entropy in any values of
$\beta$ as,
\begin{equation}\label{m5}
S(\beta)=\ln Z(\beta)+\beta E,
  \end{equation}
where $Z(\beta)$ is the partition function in which $k_{\beta} =1$ is used. By expanding $S(\beta)$ around
equilibrium temperature $(\beta =\beta_0)$ one can obtain,
\begin{equation}\label{m6}
S(\beta)= {S_0} +\frac{1}{2} (\beta - \beta_0)^2 S{''}_0+ {...},
  \end{equation}
where $S{''}(\beta_0)={(\frac{\partial
S(\beta)}{\partial\beta})}_{(\beta = \beta_0 )}=0$,
$\beta_0=\frac{1}{T_0}$, and $S(\beta_0)=S_0$. Using $S=\ln\rho(E)$ we
have [31],
\begin{equation}\label{m7}
S = S_0 -\frac{1}{2} \ln{S{''}_0}+...
  \end{equation}
From the entropy at any temperature $S(\beta)$ there is the right
canonical entropy at equilibrium temperature (7). Second
derivative of (5) gives the following relation,
\begin{equation}\label{m8}
S{''}(\beta)=\frac{1}{Z}\frac{\partial^2
S(\beta)}{(\partial\beta)^2}-\frac{1}{Z^2}\frac{\partial
S(\beta)}{(\partial\beta)^2},
  \end{equation}
where we used $<E>=-(\frac{\partial S(\beta)}{\partial
\beta})_{(\beta=\beta_0)}$.\\
We have $S{''}(\beta)$ as a fluctuation square of energy. As
$C=({\frac{\partial E}{\partial T}})_{(T=T_0 )}$  the following
equation will be obtained,
\begin{equation}\label{m9}
S{''}(\beta)=CT^2
  \end{equation}
By putting (9) in (7) the corrected entropy will be,
  \begin{equation}\label{m10}
S=S_0-\frac{1}{2}\ln{CT^2 }+...,
  \end{equation}
We will use this formula for some black holes in next section. For
black holes we have $T_0 = T_H$. If we consider $16\pi G_N =1$, therefore,
\begin{equation}\label{m11}
S_0 = S_{BH}=4\pi r_+.
  \end{equation}
We need some thermodynamic properties of black holes like,
\begin{equation}\label{m12}
T=\frac{dM}{dS},
  \end{equation}
and,
 \begin{equation}\label{m13}
C=\frac{dM}{dT}=T \frac{dS}{dT},
  \end{equation}
where $M$ is the mass of black hole, by replacing them to
(10) the entropy will be obtained.
\subsection{Charged BTZ black hole}
In the equation (2) if $B=0$ we have a black hole without
coupling with scalar field, the metric function becomes,
\begin{equation}\label{m14}
f(r)=\frac{r^2}{l^2} - M - \frac{Q^2}{2}\ln{(r)},
  \end{equation}
  therefore the black hole mass establish as the following,
\begin{equation}\label{m15}
M=\frac{{r_+}^2}{l^2} - \frac{Q^2}{2}\ln{(r_+)}.
  \end{equation}
By setting $r_+$ in terms of $S_{BH}$, we have,
\begin{equation}\label{m16}
M=-\frac{Q^2}{2}\ln{\frac{S_0}{4\pi}} + \frac{{S_0}^2}{16\pi^2 l^2}.
  \end{equation}
Using  equation (15)  the temperature is given by,
\begin{equation}\label{m17}
T= -\frac{Q^2}{2S_0} + \frac{S_0}{8\pi^2 l^2}
  \end{equation}
From (13) and (17) we can calculate $C$, putting on
(10), we obtain,
\begin{equation}\label{m18}
S= S_0 + \frac{1}{2}\ln {S_0} + \frac{1}{2}\ln({{S_0}^2 + 4\pi^2 l^2
Q^2}) - \frac{3}{2}\ln({{S_0}^2 - 4\pi^2 l^2 Q^2}) + \ln{8\pi^2l^2}.
  \end{equation}
If we set $Q=0$ we have BTZ black hole and the logarithmic
correction as the following,
\begin{equation}\label{m19}
S= S_0-\frac{1}{2}\ln{S_0}+\ln{8\pi^2l^2}
  \end{equation}
This is the same result as we saw in [31].
\subsection{Uncharged hairy AdS black hole}
In this case we have,
\begin{equation}\label{m20}
f(r)=-M(1 + \frac{2B}{3r}) + \frac{r^2}{l^2},
  \end{equation}
and therefore,
\begin{equation}\label{m21}
M=\frac{{r_+}^2}{l^2(1 + {\frac{2B}{3r_+}})}.
  \end{equation}
So the following term will be the temperature,
\begin{equation}\label{m22}
T=\frac{9{S_0 }^2(S_0 + 4\pi B)}{8\pi^2 l^2({3S_0 + 8\pi B})^2}.
  \end{equation}
The correct entropy can be,
\begin{equation}\label{m23}
S= S_0 - \frac{5}{2}\ln{S_0} +\frac{3}{2}\ln{\frac{3S_0 + 8\pi
B}{S_0 +4\pi B}} + \frac{1}{2}\ln({3{S_0 }^2 + 24\pi BS_0 + 64\pi^2
B^2})+\ln{\frac{8\pi^2 l^2}{9}}.
  \end{equation}
Therefore if the coupling of scalar field be ignored we have,
\begin{equation}\label{m24}
S= S_0 - \frac{3}{2}\ln{S_0} + \ln{8\pi^2 l^2}.
  \end{equation}
Again the leading order correction is logarithmic.
\subsection{Conformally dressed black hole}
Another black hole which gives $\beta = -\frac{B^2}{l^2}$, has the
following metric function,
\begin{equation}\label{m25}
f(r)= -3\frac{B^2}{l^2} - \frac{2B^3}{{l^2}{r}} + \frac{r^2}{l^2}.
  \end{equation}
Because of $\beta=-\frac{M}{3}$ the mass will be,
\begin{equation}\label{m26}
M=\frac{{3S_0 }^2}{64{\pi^2} {l^2}}
  \end{equation}
So the temperature and specific heat are,
\begin{equation}\label{m27}
T=\frac{3S_0 }{32\pi^2l^2},
  \end{equation}
and
\begin{equation}\label{m28}
C= S_0.
  \end{equation}
Therefore,
\begin{equation}\label{m29}
S= S_0-\frac{3}{2}\ln{S_0}+\ln{\frac{32\pi^2l^2}{3}}.
  \end{equation}
As it can be seen, this is the logarithmic that we predicted before,
now we want to study the charged black hole which coupled to scalar
field.
\subsection{Charged hairy black hole in three dimensional}
By using the explanation of $\beta$ in (4), the following
metric function obtained,
\begin{equation}\label{m30}
f(r)=\frac{r^2}{l^2}-M-\frac{2MB}{3r}-\frac{Q^2 B}{18r}-Q^2
({\frac{1}{2}+\frac{B}{3r}})\ln(r).
  \end{equation}
If we set $f(r_+ )=0$ the mass will be reduced to the following equation,
\begin{equation}\label{m31}
M=\frac{9{S_0 }^3+32\pi^3 l^2 Q^2
B-72\pi^2 l^2 Q^2 S_0 \ln(\frac{S_0 }{4\pi}) - 192\pi^3 l^2 Q^2 B
\ln{\frac{S_0}{4\pi}}}{48\pi^2 l^2 (3S_0 +8\pi B)}.
  \end{equation}
Applying the equations (12) and (13) to the equation (31) we yield to the following temperature,
\begin{equation}\label{m32}
T=\frac{{9{S_0 }^4+36\pi B{S_0
}^3-36\pi^2 l^2 Q^2 {S_0}^2-208\pi^3 l^2 Q^2 BS_0 - 256\pi^4 l^2 Q^2
B^2}}{8\pi^2 l^2 ({3S_0+8\pi B})^2}.
  \end{equation}
The related leading order correction for this black hole get
\begin{eqnarray}\label{m33}
S=S_0 &+& {\ln(S_0)} + 2{\ln(3S_0 + 8\pi B)} + \frac{1}{2}
\ln(27{S_0}^3+ 72\pi B{S_0}^2  - 36\pi^2 l^2 Q^2 S_0 +
\frac{256\pi^4{l^2}{Q^2}{B^2}}{S_0} \nonumber\\
&-&\frac{6}{3S_0+8\pi
B}(9{S_0}^4 + 36\pi B{S_0}^3 - 3\pi^2 {l^2} {Q^2}{S_0}^2 - 208\pi^3
l^2 Q^2 BS_0 - 256\pi^4 l^2 Q^2 B^2)) \nonumber\\
&-&\frac{3}{2}\ln(9{S_0}^4 + 36\pi B{S_0}^3 - 36\pi^2 {l^2}
{Q^2}{S_0}^2 - 208 \pi^3 {l^2} {Q^2} S_0 - 256\pi^4 {l^2} {Q^2}
{B^2})\nonumber\\
&+& \ln(8\pi^2 {l^2}).
  \end{eqnarray}
If the scalar field be discarded $(B=0)$, we have the
charged BTZ black hole and we see the same entropy as equation
(19).\\
In the equation (32) when we neglect charge $(Q=0)$ we have the
similar result of (19).\\
We apply $Q=0$ and $B=0$ on (33) to obtain,
\begin{equation}\label{m34}
S= S_0-\frac{3}{2}\ln{S_0}+\ln{8\pi^2 l^2}
  \end{equation}
In next step we show the exact of these special black holes by the suggestion form of CFT.

\section{Exact Entropy Function}
In previous stages we have seen that the logarithmic corrections can
be obtained from thermodynamics properties like temperature and
specific heat of black holes. we will try to make over these
corrections by considering an exact entropy function $S(\beta)$,
which follows from $CFT$ and other quantum theories of gravity. If
we return to the equation (7) for using it to study entropy of some black
holes we need $S{''}(\beta_0)$. By considering the special form of
function of entropy that is approximated by the parabolic form
(6). The special form is,
\begin{equation}\label{m35}
S(\beta)=x\beta+\frac{y}{\beta}.
  \end{equation}
A general form of entropy function was assumed to check necessity and influenced on results, that is,
\begin{equation}\label{m36}
S(\beta)=x\beta^m+\frac{y}{\beta^n},
  \end{equation}
where the case $m=n=1$ is commanded by $CFT$. By finding the
extremum of (\ref{m35}) and getting the second derivative of
$S(\beta)$ then by some calculation similar to the Ref. [31] we
have the following form of entropy,
\begin{equation}\label{m37}
S=S_0-\frac{1}{2}\ln{S_0 T^2}.
\end{equation}
Now we use the above entropy and apply it to previous black holes.
\subsection{Charged BTZ black hole}
In that case we obtain,
\begin{equation}\label{m38}
S= S_0 + \frac{1}{2}\ln{S_0} - \ln{{S_0}^2 - 4\pi^2 l^2 Q^2} +
\ln{8\pi^2 l^2}.
  \end{equation}
We see that this is same as (18) by one different expression that
is,
\begin{equation}\label{m39}
\frac{1}{2}\ln{\frac{{S_0}^2 + 4\pi^2 l^2 Q^2}{{S_0}^2 - 4\pi^2 l^2
Q^2}},
  \end{equation}
so if $Q=0$ the logarithmic correction of BTZ black hole will
obtained.
\subsection{Uncharged hairy AdS black hole}
By using (37) for this black hole the entropy will be,
\begin{equation}\label{m40}
S= S_0 - \frac{5}{2}\ln{S_0} + \ln{\frac{(3S_0+8\pi B)^2}{S_0 + 4\pi
B}} + \ln{\frac{8\pi^2 l^2}{9}}.
  \end{equation}
There is one expression that makes it different from (23),
this is,
\begin{equation}\label{m41}
 \frac{1}{2}\ln{\frac{3{S_0 }^2 + 24\pi BS_0 + 64\pi^2
B^2}{{3S_0 +8\pi B}{S_0 + 4\pi B}}},
  \end{equation}
if we set $B=0$, we see the
same relation as (19).
\subsection{Conformally dressed black hole}
For this part the answer is the same as (29) as we
predicted.
\subsection{Charged hairy black hole}
In that case we calculate,
\begin{eqnarray}\label{m42}
S=S_0&+&\frac{1}{2}\ln{S_0} + 2\ln(3S_0 + 8\pi B)-\ln(9{S_0}^4 + 36\pi
B{S_0}^3-36\pi^2 l^2 Q^2{S_0 }^2 \nonumber\\
&-& 208\pi^3 l^2 Q^2 BS_0
- 256\pi^4 l^2 Q^2 B^2) + \ln{8\pi^2 l^2}.
  \end{eqnarray}
This is the leading order correction of (33) without two
expressions that is,
\begin{align}\label{m43}
\frac{1}{2}[\ln(27(S_0)^4 + 72\pi B{S_0}^3 - 36\pi^2 l^2 Q^2 {S_0}^2
+ 256\pi^4 l^2 Q^2 B^2 - \frac{6S_0}{3S_0+8\pi B}(9{S_0}^4
\nonumber\\ + 36\pi B{S_0}^3 -36\pi^2 l^2 Q^2 {S_0}^2 - 208\pi^3 l^2
Q^2 BS_0 - 256\pi^4 l^2 Q^2 B^2)) \nonumber\\ - \ln({9{S_0}^4 +
36\pi B{S_0}^3 - 36\pi^2 l^2 Q^2 {S_0}^2 - 208\pi^3 l^2 Q^2 B S_0 -
256\pi^4 l^2 Q^2 B^2})],
  \end{align}
and if the scalar field and
charge are ignored we will derive the same shape of (29).
We saw that both of these ways for getting the corrections have the same result as shows above.
\section{Conclusion}
In this paper, we studied some special types of charged black
hole in three dimensions coupled to the scalar fields. We calculated
the logarithmic correction for these black holes with two
approaches. The first approach is the thermodynamics properties
and the second one is the entropy function, which is suggested by
CFT area. We obtained corrections of the entropy as the natural
logarithm of the cosmological constant for four different kinds of
black hole and found good agreement between them. This means
that the presence of the cosmological constant is necessary to
have logarithmic corrections. The completed form of the black
hole led to (38) using the first method and (53) using the second
method, where logarithmic corrections were illustrated. It is obvious
that the logarithmic corrections increase entropy and yield
more radiation from the black hole. These logarithmic corrections
are due to thermal fluctuations of the black hole around its
equilibrium. In this paper we achieved the logarithmic form of
entropy. We compared the results and demonstrated that by two
approaches. Here we have seen that the results for the two approaches
are the same without the scalar fields and the electric
charge. In this case the corresponding black hole will be a BTZ
black hole.


\begin{thebibliography}{11}
\bibitem{1} Deser, S., Jackiw, R., and Templeton, S. (1982). Topologically massive gauge theories. \emph{Annals of Physics, 140(2)}, 372-411.

\bibitem{2} Deser, S., Jackiw, R., and Templeton, S. (1982). Three-dimensional massive gauge theories. \emph{Physical Review Letters, 48(15)}, 975.
\bibitem{3} Deser, S., and Jackiw, R. (1984). Three-dimensional cosmological gravity: dynamics of constant curvature. \emph{Annals of Physics, 153(2)}, 405-416.

\bibitem{4} Giddings, S., Abbott, J., and Kuchar, K. (1984). Einstein's theory in a three-dimensional space-time. \emph{General Relativity and Gravitation, 16(8)}, 751-775.

\bibitem{5} Banados, M., Teitelboim, C., and Zanelli, J. (1992). Black hole in three-dimensional spacetime. \emph{Physical Review Letters, 69(13)}, 1849.

\bibitem{6} Martinez, C., Teitelboim, C., and Zanelli, J. (2000). Charged rotating black hole in three spacetime dimensions. \emph{Physical Review D, 61(10)}, 104013.
\bibitem{7} Xu, W., and Zhao, L. (2013). Charged black hole with a scalar hair in $(2+ 1)$ dimensions. \emph{Physical Review D, 87(12),} 124008.
\bibitem{P8}
L. Zhao, W. Xu, B. Zhu, "Novel rotating hairy black hole in
(2+1)-dimensions", [arXiv:1305.6001 [gr-qc]]
\bibitem{P9}
J. Sadeghi, B. Pourhassan, H. Farahani, "Rotating charged hairy black hole in (2+1) dimensions and particle acceleration", [arXiv:1310.7142 [hep-th]]
\bibitem{P10}
A. Belhaj, M. Chabab, H. EL Moumni, M. B. Sedra, "Critical Behaviors
of 3D Black Holes with a Scalar Hair", [arXiv:1306.2518 [hep-th]]
\bibitem{P11}
J. Sadeghi, H. Farahani, "Thermodynamics of a charged hairy black
hole in (2+1) dimensions", [arXiv:1308.1054 [hep-th]]
\bibitem{12} Bekenstein, J. D. (1973). Black holes and entropy. \emph{Physical Review D, 7(8)}, 2333.; Hawking, S. W. (1976). Black holes and thermodynamics.\emph{ Physical Review D, 13(2)}, 191.
\bibitem{13} Heckler, A. F. (1997). Formation of a Hawking-radiation photosphere around microscopic black holes. \emph{Physical Review D, 55(2)}, 480.
\bibitem{14} Das, S., Majumdar, P., and Bhaduri, R. K. (2002). General logarithmic corrections to black-hole entropy. \emph{Classical and Quantum Gravity, 19(9)}, 2355.
\bibitem{15} Natsuume, M., Okamura, T., and Sato, M. (2000). Three-dimensional gravity with a conformal scalar field and asymptotic Virasoro algebra. \emph{Physical Review D, 61(10)}, 104005.
\bibitem{16} Correa, F., Martinez, C., and Troncoso, R. (2011). Scalar solitons and the microscopic entropy of hairy black holes in three dimensions. \emph{Journal of High Energy Physics, 2011(1)}, 1-15.
\bibitem{17} Correa, F., Martinez, C., and Troncoso, R. (2012). Hairy black hole entropy and the role of solitons in three dimensions. \emph{Journal of High Energy Physics, 2012(2)}, 1-21.
\bibitem{18} Zeng, D. F. (2009). An exact hairy black hole solution for AdS/CFT superconductors. \emph{arXiv preprint arXiv:0903.2620}.
\bibitem{19} Myung, Y. S. (2008). Phase transition for black holes with scalar hair and topological black holes. \emph{Physics Letters B, 663(1)}, 111-117.
\bibitem{20} Bekenstein, J. D. (1996). \emph{Black hole hair} (No. gr-qc/9605059).
\bibitem{21} Abramo, L. R., Brenig, L., Gunzig, E., and Saa, A. (2003). A note on dualities in Einstein's gravity in the presence of a non-minimally coupled scalar field. \emph{Modern Physics Letters A, 18(15)}, 1043-1051.
\bibitem{22} Acena, A., Anabalon, A., and Astefanesei, D. (2012). Exact hairy black brane solutions in $ AdS_ {5} $ and holographic RG flows. \emph{arXiv preprint arXiv:1211.6126}.
\bibitem{23} Aparicio, J., Grumiller, D., Lopez, E., Papadimitriou, I., and2 Stricker, S. (2013). Bootstrapping gravity solutions. \emph{Journal of High Energy Physics, 2013(5)}, 1-48.
\bibitem{24} Dias, O. J., Horowitz, G. T., and Santos, J. E. (2011). Black holes with only one Killing field. \emph{Journal of High Energy Physics, 2011(7)}, 1-43.
\bibitem{25} Degura, Y., Sakamoto, K., and Shiraishi, K. (2001). Black Holes with Scalar Hair in $(2+1)$ dimensions. \emph{Grav. Cosmol, 7}, 153-158.
\bibitem{26} Banados, M., and Theisen, S. (2005). \emph{Scale invariant hairy black holes} (No. hep-th/0506025).
\bibitem{27} Kolyvaris, T., Koutsoumbas, G., Papantonopoulos, E., and Siopsis, G. (2011). Einstein hair. \emph{arXiv preprint arXiv:1111.0263}.
\bibitem{28} Brihaye, Y., and Hartmann, B. (2012). A scalar field instability of rotating and charged black holes in $(4+1)$-dimensional Anti-de Sitter space-time. \emph{Journal of High Energy Physics, 2012(3)}, 1-22.
\bibitem{29} Anabalon, A., and Maeda, H. (2010). New charged black holes with conformal scalar hair. \emph{Physical Review D, 81(4)}, 041501.
\bibitem{30} Nucamendi, U., and Salgado, M. (2003). Scalar hairy black holes and solitons in asymptotically flat spacetimes. \emph{Physical Review D, 68(4)}, 044026.
\bibitem{31} Das, S., Majumdar, P., and Bhaduri, R. K. (2002). General logarithmic corrections to black-hole entropy. \emph{Classical and Quantum Gravity, 19(9)}, 2355.

\end{thebibliography}
\end{document}